\documentclass[useAMS,usenatbib]{mn2e}
\usepackage{times, graphicx}

\title[Tracking Magnetic Bright Point Motions]
  {Tracking Magnetic Bright Point Motions Through the Solar Atmosphere}
\author[P.~H.~Keys et al.]
  {P.~H.~Keys$^1$\thanks{Email: pkeys02@qub.ac.uk}, M.~Mathioudakis$^1$, D.~B.~Jess$^1$, 
S.~Shelyag$^1$, \newauthor D.~J.~Christian$^2$ and F.~P.~Keenan$^1$\\
  $^1$Astrophysics Research Centre, School of Mathematics and Physics, Queen's University, Belfast, BT7~1NN, Northern Ireland, U.K.\\
  $^2$Department of Physics and Astronomy, California State University, Northridge, CA 91330, USA}

\begin{document}
\date{Accepted 2012 October 22. Received 2012 October 15; in original form 2012 August 10}
\pagerange{\pageref{firstpage}--\pageref{lastpage}} \pubyear{2012}
\maketitle
\label{firstpage}


\begin{abstract}
High cadence, multi-wavelength observations and simulations are employed 
for the analysis of solar photospheric magnetic bright points (MBPs) in the quiet Sun. 
The observations were obtained with the Rapid Oscillations in the Solar Atmosphere ({\it{ROSA}})
imager and the Interferometric BIdimensional Spectrometer ({\it{IBIS}}) at the Dunn Solar 
Telescope. Our analysis reveals that photospheric MBPs have an average 
transverse velocity of approximately 1~km\,s$^{-1}$, whereas their chromospheric 
counterparts have a slightly higher average velocity of 1.4~km\,s$^{-1}$. Additionally, chromospheric 
MBPs were found to be around 63\% larger than the equivalent photospheric 
MBPs. These velocity values were compared with the output of numerical simulations 
generated using the MURaM code. The simulated results were similar, but slightly 
elevated, when compared to the observed data. An average velocity of 
1.3~km\,s$^{-1}$ was found in the simulated G-band images and an 
average of 1.8~km\,s$^{-1}$ seen in the velocity domain at a height 
of 500~km above the continuum formation layer. Delays in the change of velocities were 
also analysed. Average delays of $\sim$4~s between layers of the simulated data set were established 
and values of $\sim$29~s observed between G-band and Ca~{\sc{ii}}~K {\it{ROSA}} observations. 
The delays in the simulations are likely to be the result of oblique granular shock waves, whereas those found 
in the observations are possibly the result of a semi-rigid flux tube.
\end{abstract}


\section{Introduction}
\label{Intro}
Magnetic bright points (MBPs) are ubiquitous in the solar photosphere  \citep{Dunn73}.
They are thought to be the footpoints of magnetic flux tubes in the inter-granular lanes with field strengths of the order of a kilogauss \citep{Sten85, Solan93}. Their brightness is due to the reduced pressure within the flux tube allowing the observer to view a 
deeper, hotter region of the photosphere, and is also due to the heating of the plasma within the flux tube by material surrounding its walls. The  higher temperature reduces the abundance of the CH molecule, making MBPs appear brighter in G-band ($4305$~{\AA}) images  \citep{Stein01, Shel04}. G-band observations 
therefore provide a good base for analysing the photospheric properties of these concentrated magnetic structures.

The importance of MBP studies was reviewed in \citet{deWi09}. Briefly, studies of MBP motions are 
important as these may drive Alfv\'{e}n waves in magnetic field lines within flux tubes,
where the energy carried by the waves can be transmitted to the corona and lead to heating. This has been
a recent topic of several 3-D MHD models \citep{vBall11, Asgari12}
investigating these effects.

Recent MBP studies have focussed on their size, velocity distributions  \citep{San04, Croc10, Keys2011} and intensity peaks \citep{Lan02}.  \citet{Bon08} have identified convective downdrafts and vortex motions at the location of MBPs, phenomena that have also been reproduced in simulations of radiative MHD \citep{Shel11a,Shel11b,Shel12}.


A study of MBPs in the Na{\,}{\sc{i}}{\,}D{$_1$} line  \citep{Jess10a} has shown evidence for strong downdrafts with velocities as high at 7~km{\,}s$^{-1}$ at MBP centres. 
It was also noted that field strengths affected the expansion 
of the flux tubes, with weak fields leading to an expansion of $\approx$76\% whereas higher field 
strengths led to an expansion of $\approx$44\%.

\begin{figure*}
\centering
   \includegraphics[width=16.5cm]{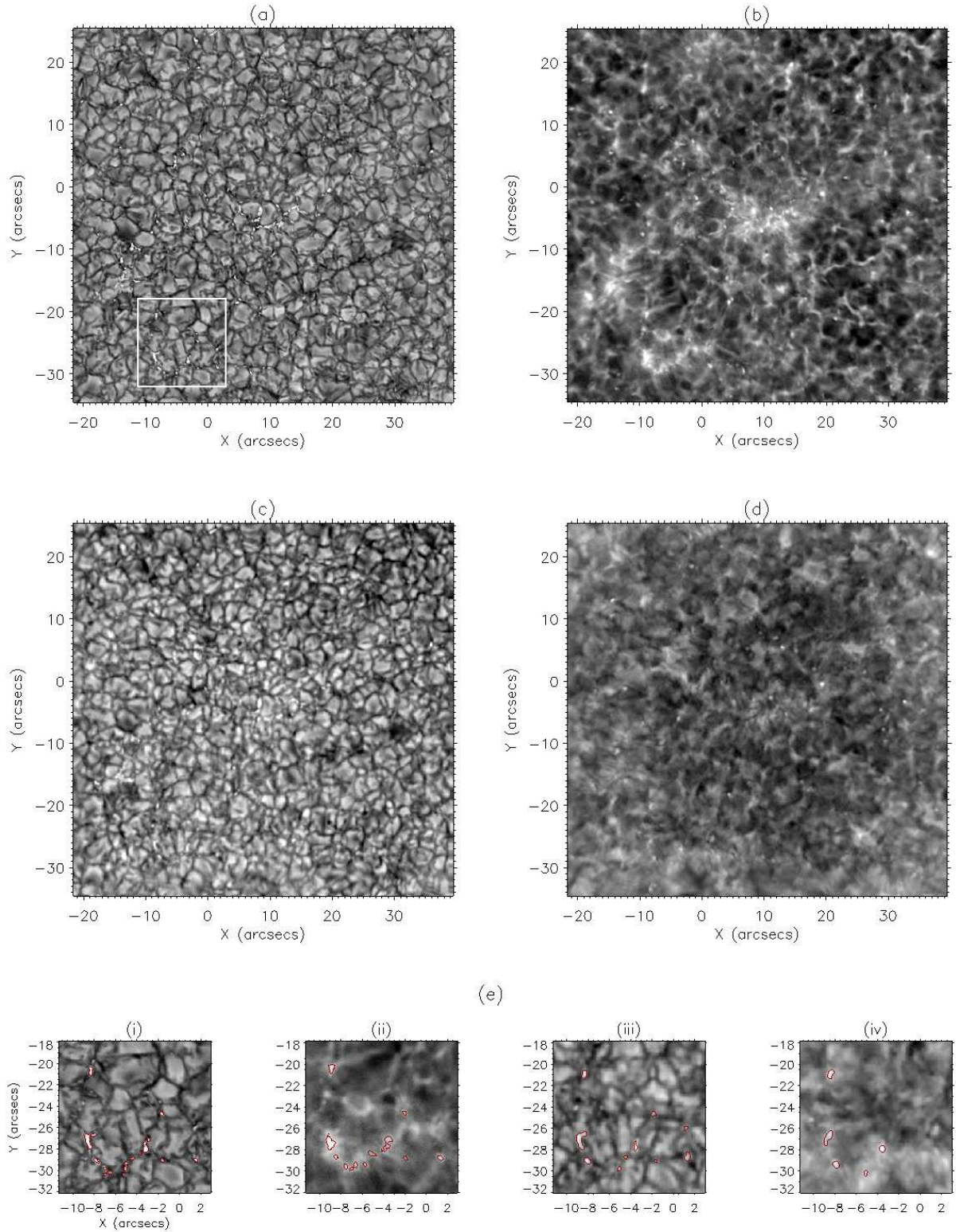}
     \caption{Sample {\it{ROSA}} and {\it{IBIS}} data employed in this study. 
		\textbf{(a)} G-band image \textbf{(b)} Co-spatial and co-temporal 
		image in Ca\,{\sc{ii}}~K. \textbf{(c)} and \textbf{(d)} Co-spatial {\it{IBIS}} images in the wing and the core 
		of the Na~{\sc{i}}~D{$_1$} line, respectively. MBPs are abundant in \textbf{(a)}. However, higher in the atmosphere their boundaries 
		become more ambiguous due to reverse granulation.\textbf{(e)} An expanded view of the 4 data sets in the region outlined 
		by the \textit{white} box in \textbf{(a)}. \textbf{(i)},  \textbf{(ii)},  \textbf{(iii)} and  \textbf{(iv)} show expanded regions in G-band, Ca~{\sc{ii}}~K 
		and the wing and core of the Na~{\sc{i}}~D{$_1$} line, respectively. A small sample of MBPs are contoured \textit{red} in the 4 data sets showing 
		both groups of MBPs and isolated MBPs. The different numbers and sizes of MBPs is evident across the 4 data sets.
     }
     \label{Fig1}
\end{figure*}

\citet{Keys2011} studied the transverse velocities of over 6000 MBPs in both observed and simulated G-band 
images. 
Their study revealed an average velocity of 1~km{\,}s$^{-1}$ and a maximum value of 7~km{\,}s$^{-1}$. Radiative MHD simulations 
with an average field strengths of 200~G and 400~G revealed similar velocity distributions, the latter showing the 
best agreement with observations. 

\citet{Ulm91} have shown that the horizontal velocity of solar magnetic flux tubes increases as a function of height. The amplitude of the horizontal velocity component scales as $\rho^{-1/2}$, where $\rho$ is the density, and the authors note that the transverse wave flux is only approximately conserved. In the regime of strict flux conservation, a horizontal velocity 
component of 1.5~km{\,}s$^{-1}$ was expected at a height of 340~km, yet a velocity of only 0.8~km{\,}s$^{-1}$ was observed. This is very similar to the values of  \citet{Keys2011}.  

\begin{figure*}
\centering
   \includegraphics[width=16.5cm]{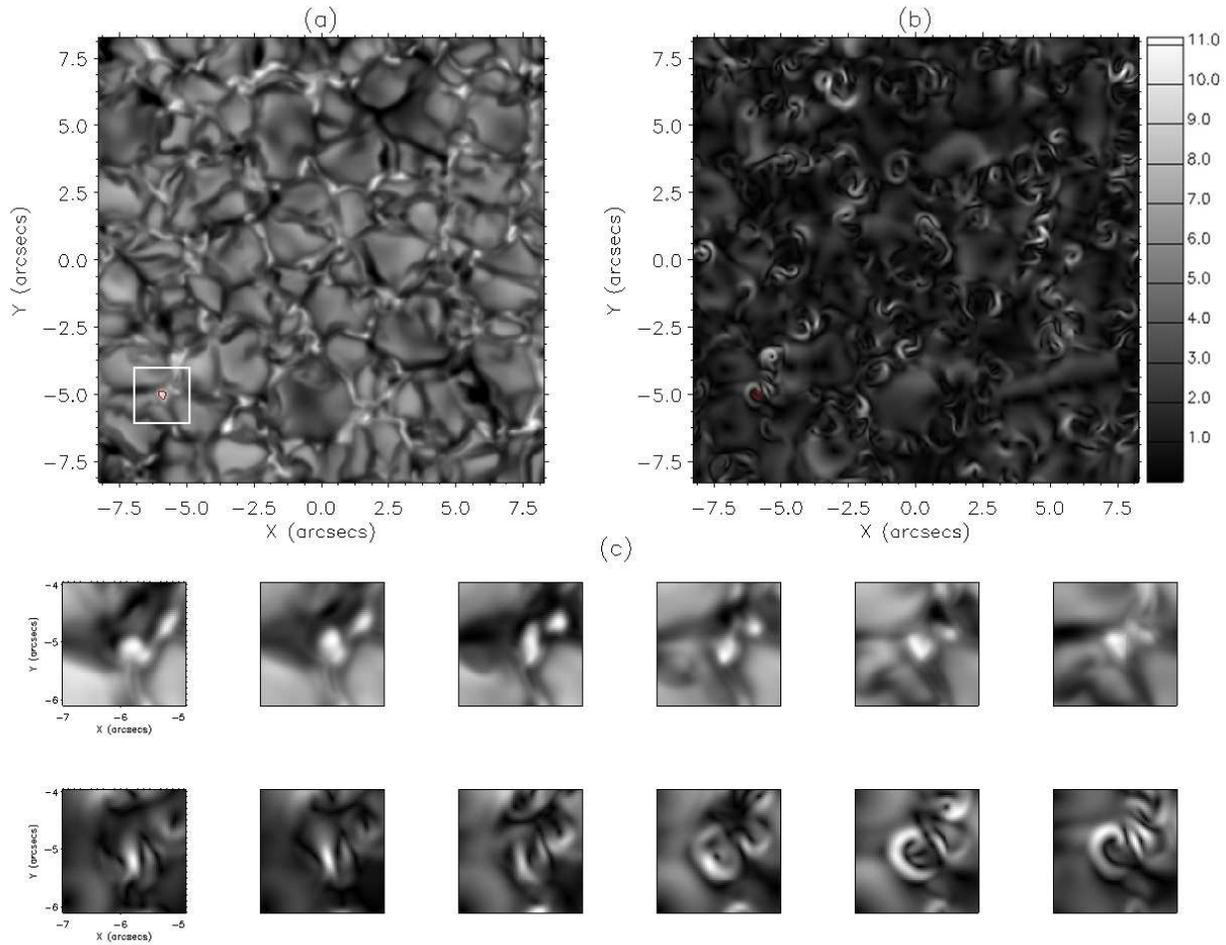}
     \caption{Sample images from the simulated data sets. \textbf{(a)} Snapshot of a 12$\times$12~Mm G-band 
		simulation. \textbf{(b)} The corresponding velocity domain 500~km above the continuum formation layer for the image shown in \textbf{(a)}. 
		Velocity scale is in km\,s$^{-1}$. A sample, isolated MBP, is contoured \textit{red} in  \textbf{(a)} and 
		\textbf{(b)}. \textbf{(c)} shows a set of expanded panels depicting the evolution of a MBP found in the \textit{white} box in \textbf{(a)}. 
		The top row shows the G-band images of the MBP advancing in time by $\sim$44~s between each panel from left to right. The bottom 
		row shows the corresponding velocity domain images for the MBP. The final panel shows the MBP as seen in \textbf{(a)} and \textbf{(b)}.
     }
     \label{Fig2}
\end{figure*}

MBPs have average diameters of around 135~km \citep{San04} and lifetimes of approximately 90~s \citep{Keys2011}, and hence tracking these throughout the solar atmosphere requires  high spatial and temporal resolution imaging. The accurate co-alignment of the images is vital when multi-wavelength studies are considered. In this 
paper we use simultaneous multi-wavelength observations to study MBPs at various heights in the solar atmosphere and improved BP tracking 
algorithms over our previous studies \citep{Croc10}. We present  multi-wavelength observations and numerical 
simulations of photospheric MBPs in co-temporal and co-spatial G-band, Ca~{\sc{ii}}~K  and Na~{\sc{i}}~D$_1$ images. 
The observations and numerical simulations used are described in Section~\ref{obs}. Analysis of these results as well as their 
implications are presented in Section~\ref{anal}, with concluding remarks given in Section~\ref{conc}.

\section{Observations and Numerical Simulations}
\label{obs}
The imaging data employed in this study were obtained using the Rapid Oscillations in the Solar Atmosphere 
\citep[{\it{ROSA}};][]{Jess10b} instrument, at the Dunn Solar Telescope (DST), in New Mexico, USA. Observations were obtained during a 
period of excellent seeing on 2009 May 28, using a $9.2$~{\AA} wide filter 
centred at $4305$~{\AA} (G-band) and a $1$~{\AA} wide filter centred at $3933.7$~{\AA} (Ca~{\sc{ii}}~K core). 
We observed a $69{\arcsec} \times 69{\arcsec}$ quiet Sun region at disk centre 
for $\sim$50~minutes, using a common plate scale of $0{\arcsec}.069$~pixel$^{-1}$ for the {\it{ROSA}} cameras.
The images were reconstructed using the Speckle algorithm of \citet{Wog08}, while image de-stretching 
was performed using a $40 \times 40$ grid \citep[equating to a $\approx$$1{\arcsec}.7$ separation between 
spatial samples;][]{Jess08}.  G-band images were taken at a raw 
cadence of 0.033~s, while after reconstruction the cadence was reduced to 0.528~s. 
Reconstructed G-band images were then binned into consecutive groups of four to improve the 
signal-to-noise and reduce the overall volume of the dataset, providing a final 
image cadence of 2.1~s. The Ca~{\sc{ii}}~K images had a raw cadence of 0.26~s, and a post-speckle 
reconstructed cadence of 4.2~s. Subsequently, the two data sets were spatially co-aligned.

The Interferometric BIdimensional Spectrometer \citep[{\it{IBIS}};][]{Cav06} was employed in conjunction 
with {\it{ROSA}}, covering the Na~{\sc{i}}~D{$_1$} absorption line 
at 5895.94~{\AA}. At a spatial sampling of $0{\arcsec}.097$~pixel$^{-1}$, the near square field-of-view of 
{\it{ROSA}} was contained within the circular aperture of {\it{IBIS}}. Nine wavelength steps were utilised with {\it{IBIS}},
with 10 images per wavelength step to aid image reconstruction, resulting in a complete scan cadence 
of 39.7~s. A blueshift correction was required due to the classical etalon mountings 
\citep{Cauz08}. Images of the Na~{\sc{i}}~D{$_1$} core used for MBP studies are 
true intensity maps, created by establishing the line-profile minimum at each 
pixel. By displaying Doppler-compensated line-centre intensities, 
rather than rest-wavelength intensities, brightness variations 
throughout the image are more indicative of the source function than of the 
velocities present in the line-forming region \citep{Jess10a}. 
A sample of {\it{ROSA}} and {\it{IBIS}} images is shown in Figure~\ref{Fig1}.

The radiative magneto-hydrodynamic (MHD) code MuRAM \citep{Vog05}, has been used to produce 
simulated G-band images, and their corresponding velocity domain. This code solves large-eddy 
radiative three-dimensional MHD equations on a Cartesian grid, and employs a fourth-order 
Runge-Kutta scheme to advance the numerical solution in time. The numerical domain has a physical 
size of 12$\times$12~Mm$^2$ in the horizontal direction, 1.4~Mm in the vertical direction, and is resolved 
by 480$\times$480$\times$100 grid cells, respectively. Our starting point for the simulations is a 
well-developed non-magnetic ($B=0$) snapshot of photospheric convection taken approximately 2000~s 
(about 8 convective turnover timescales) from the initial plane-parallel model. A uniform magnetic field 
of 200~G was introduced at this stage, and a sequence of 829 snapshots recorded, each separated by a 
time interval of $\sim$2~s. The resulting sequence, which is used for further G-band radiative diagnostics, 
covers approximately 30~minutes of physical time, corresponding to $\sim$$4-6$ granular lifetimes. G-band 
images were produced using the radiative diagnostics technique briefly described by \citet{Jess12b}. 
The plasma velocity field corresponding to these images was measured at a height of about 500~km 
above the continuum formation layer (see Figure~\ref{Fig2}).

\section{Analysis and Results}
\label{anal}
\citet{Keys2011} employed an automated detection and tracking algorithm to estimate the velocities of 6000 G-band MBPs. 
However, an automated detection and tracking algorithm is extremely hard to employ for 
the Ca~{\sc{ii}}~K data set. Due to the reverse granulation inherent in the Ca~{\sc{ii}}~K images, 
the application of intensity thresholding for 
the identification of MBPs proved unsatisfactory. The velocities of Ca~{\sc{ii}}~K MBPs were 
instead determined using the 
Local Correlation Tracking (LCT) technique \citep{Mat10}. This was utilised along with manual tracking 
to ensure accurate velocity estimates. As shown in the G-band analysis of \citet{Keys2011}, 
the LCT technique produces very similar results to the automated tracking algorithm.    

\begin{table*}
\caption{Summary of MBP characteristics directly compared at various heights.}           
\label{table1}      
\begin{tabular}{c c c c c}  
\hline\hline
\textbf{MBP} & \textbf{Ca~\sc{ii}~K} & \textbf{Ca~\sc{ii}~K} & \textbf{G-band} & \textbf{G-band}\\                        
\textbf{No.} &  \textbf{Vel.(km\,s$^{-1}$)} &  \textbf{Area(km$^2$)}  & \textbf{Vel.(km\,s$^{-1}$)} & \textbf{Area(km$^2$)} \\
\hline                                   
1 & 0.74 & 95,000 & 0.75 & 35{\,}000 \\
2 & 3.26 & 66,250  & 1.10 & 40{\,}000 \\
3 & 2.09 & 88,750  & 1.62 & 40{\,}000 \\
4 & 1.27 & 131,250  & 0.46 & 35{\,}000 \\
5 & 1.36 & 88,750 & 0.81 & 30{\,}000\\
6 & 1.19 & 87,500 & 0.23 & 50{\,}000\\
7 & 2.04 & 142,500  &  0.13 & 102{\,}500\\
8 & 0.82 & 113,750 &  0.51 & 52{\,}500\\
9 & 2.77 & 116,250 & 2.54 & 112{\,}500\\
10 & 0.26 & 156,250 & 0.68 & 110{\,}000\\
\hline                                            
\end{tabular}
\end{table*}

The ambiguity associated with their boundaries did not allow a one-to-one correspondence between all 
the MBPs in the G-band and in Ca~{\sc{ii}}~K. Approximately 100 MBPs were analysed in Ca~{\sc{ii}}~K 
with half of them having 
unambiguous G-band counterparts.  
The velocity and area of each MBP was subsequently evaluated. As mentioned previously, the actual 
value for the area of an MBP in Ca~{\sc{ii}}~K can be difficult to determine due to the ambiguity of the MBP 
boundaries. The error in the determination of MBP surface area was established through 
the calculation of mean absolute measurement errors, and found to be approximately $\pm$23\%. 
Difficulties in defining an accurate area for Ca~{\sc{ii}}~K MBPs can be clearly seen in Figure~\ref{Fig1}\,(b). 

\begin{figure*}
\centering
   \includegraphics[width=16.5cm]{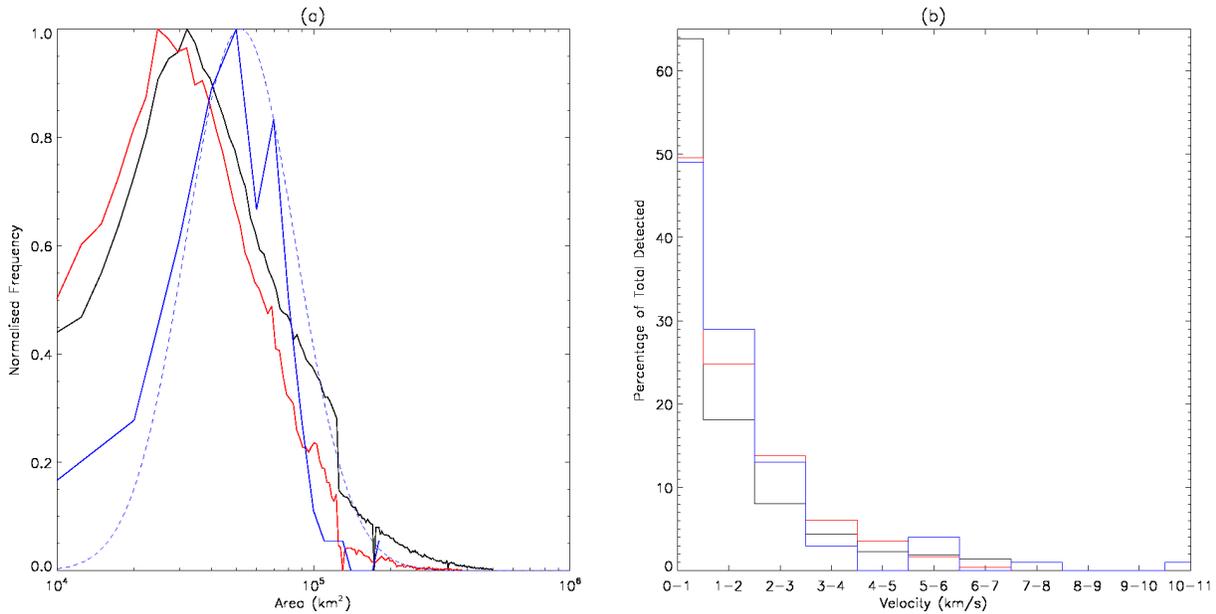}
     \caption{The area and velocity distributions of the observed and simulated datasets. \textbf{(a)} The normalised area distributions of the data sets; 
		the \textit{black} line is the observed G-band,  the \textit{red} line is the simulated G-band and the \textit{blue} line 
		is the Ca~{\sc{ii}}~K data. The log-normal distribution of the two G-band data sets is clear. However, the smaller number of 
		data points in Ca~{\sc{ii}}~K means that this distribution is not as clear. The \textit{dashed-blue} line shows a possible ideal 
		log-normal curve distribution for Ca~{\sc{ii}}~K given better count statistics. It is evident that the Ca~{\sc{ii}}~K MBPs have 
		larger areas than their G-band counterparts. \textbf{(b)} displays a histogram of the velocity 
		distributions for observed G-band (\textit{black}), simulated G-band (\textit{red}) and Ca~{\sc{ii}}~K (\textit{blue}).
		 The G-band distributions are fairly similar and that for Ca~{\sc{ii}}~K illustrates elevated MBP velocities. 
     }
     \label{Fig3}
\end{figure*}

Our sample of 100 Ca~{\sc{ii}}~K MBPs revealed an average velocity of 1.4~km\,s$^{-1}$, 
with an estimated average area of 52{\,}000~km$^2$. 
The area of Ca~{\sc{ii}}~K MBPs follows a log-normal distribution similar to the 
G-band \citep{Croc10}, with the Ca~{\sc{ii}}~K area distribution shifted to higher values 
with a broader peak 
(Figure~\ref{Fig3}\,a). Velocity distributions for the observed and simulated G-band 
datasets and Ca~{\sc{ii}}~K are shown in Figure~\ref{Fig3} (b).
The transverse velocities of Ca~{\sc{ii}}~K MBPs are enhanced compared  to the G-band 
(an average of 1~km\,s$^{-1}$ for 6236 MBPs), with the mean absolute measurement error in both 
velocity values estimated to be 0.5~km\,s$^{-1}$. Their size also appear to be larger 
by about 63\%. 
The distributions were analysed in more detail by selecting 50 Ca~{\sc{ii}}~K MBPs 
and their corresponding G-band counterparts. Again, the average transverse 
velocities are 1.8~km\,s$^{-1}$ and 0.8~km\,s$^{-1}$ in the Ca~{\sc{ii}}~K and G-band, respectively. 
The corresponding average sizes are 51{\,}800~km$^2$ and 31{\,}300~km$^2$. 
This shows a 65\% increase in the size of the MBPs as they extend upwards, in 
agreement with \citet{Jess10a}, who found an expansion 
of $\sim$70\% in their study of MBPs manifesting in the Na~{\sc{i}}~D{$_1$} line core. 
Of these 50 MBPs, $\sim$80\% are isolated and the remiander are groups of MBPs. On closer inspection the isolated MBPs
have higher average velocities in both G-band and Ca~{\sc{ii}}~K (0.8~km\,s$^{-1}$ and 1.9~km\,s$^{-1}$ respectively) compared to the 
groups of MBPs (0.6~km\,$^{-1}$ and 0.3~km\,s$^{-1}$ in G-band and Ca~{\sc{ii}}~K respectively). As is expected, MBPs  found in a group 
have a larger average area in both G-band and Ca~{\sc{ii}}~K compared to isolated MBPs. Also, the change in area between G-band and Ca~{\sc{ii}}~K 
is more significant for isolated MBPs ($\sim$57\% change for MBP groups whereas nearly twice as large on average for isolated MBPs.) This is 
likely the result of a higher field strength found in groups of MBPs as seen in \citet{Jess10a}.
The corresponding velocities for a sample of 10 of these MBPs are given in Table~\ref{table1}.

We have also studied the MBPs visible in the Na~{\sc{i}}~D{$_1$} line. The scanning capability of 
{\it{IBIS}} allows us to study the MBPs both in the core and wings of the line, albeit at a reduced 
temporal resolution of 39.7~s. 
This resulted in a smaller sample of 50 MBPs present in both the wing and line core 
(Figure~\ref{Fig1} (c) and (d)).
The average velocity for the core of the Na~{\sc{i}}~D{$_1$} line was found to be 
only 0.7~km\,s$^{-1}$ and the 
average for the 486 MBPs analysed at the wing of the line 0.7~km\,s$^{-1}$, with a 
maximum value of 6~km\,s$^{-1}$. This maximum value is close to that found for the G-band. 
However, there is a slight discrepancy between the other velocities established for Ca~{\sc{ii}}~K, G-band and 
Na~{\sc{i}}~D{$_1$}. This is probably a result of the combination of a smaller sample size and 
the lower temporal resolution of the Na~{\sc{i}}~D{$_1$} data sets. The errors in the velocity values in 
the core and the wing of the Na~{\sc{i}}~D{$_1$} line are estimated to be 0.6~km\,s$^{-1}$ and 0.7~km\,s$^{-1}$ respectively.

The observational results were compared with radiative MHD simulations produced with the MURaM code. These 
simulations are described in \citet{Jess12a} and have a cadence of $\approx$2.1~s. We have 724 
G-band MBPs in these 200~G simulated images, which have 
 an average velocity of 1.3~km\,s$^{-1}$, a maximum velocity of 6.9~km\,s$^{-1}$, and a mean lifetime of 
110~s. The simulated MBP velocities have an error of approximately 0.5~km\,s$^{-1}$.
A summary of results from the observed and simulated data sets is given in Table~\ref{table2}. 
\begin{table*}
\caption{Summary of MBP characteristics for observations ({\it{ROSA}}/{\it{IBIS}}) and simulations (MURaM).}           
\label{table2}      
\centering                                    
\begin{tabular}{l c c c c c c}         
\hline\hline
\textbf{Data Set}  & \textbf{FOV } & \textbf{Av.} & \textbf{No.}  & \textbf{Av. Vel.} & \textbf{Max. Vel.} & \textbf{Av. Area}\\
  & \textbf{Size} & \textbf{Cadence(s)} & \textbf{MBPs} & \textbf{(km\,s$^{-1}$)} & \textbf{(km\,s$^{-1}$)} & \textbf{(km$^2$)} \\
\hline                                   

\textbf{G-band} & $70''\times70''$ & 2.1 & 6236 & 1 & 7 & 32{\,}000 \\
\textbf{({\it{ROSA}})} \\
\\
\textbf{Ca~{\sc{ii}}~K} &$70''\times70''$ & 4.2 & 100 & 1.4 & 10.2 & 52{\,}000 \\
\textbf{({\it{ROSA}})} \\
\\
\textbf{Na~{\sc{i}}~D{$_1$}} & $83''\times83''$ & 39.7 & 486 & 0.7 & 6 & 51{\,}200 \\
\textbf{(Wing-{\it{IBIS}})} \\
\\
\textbf{Na~{\sc{i}}~D{$_1$}} & $83''\times83''$ & 39.7 & 50 & 0.7 & 2.4 & 73{\,}500\\  
\textbf{(Core-{\it{IBIS}})} \\                                       
\\
\textbf{G-band} & $16''\times16''$ &  2.1 & 724 & 1.3 & 6.9 & 26{\,}100\\ 
\textbf{(MURaM)} \\
\\
\textbf{Velocity Domain} & $16''\times16''$ & 2.1 & 724 & 1.8 & 15 & N/A\\
\textbf{(MURaM)} \\
\hline   

\end{tabular}
\end{table*}

The MURaM simulations also allow us to study the corresponding velocity field at a height of 500~km above the continuum formation layer. 
Transverse velocities could therefore be established for each point in the data set along with an inclination 
angle. The velocity maps were compared to the MBPs detected in 
the simulated G-band images and used to determine an average velocity for MBPs at this height of 
1.8~km\,s$^{-1}$, with a maximum value of 15~km\,s$^{-1}$. 

Our simulations reproduce the measured G-band and Ca~{\sc{ii}}~K values, albeit with slightly elevated velocities when 
compared to the observations. However, if we were to consider only the 50 unambiguous MBPs identified in both 
bandpasses, we can see that the velocities for MBPs found higher in the atmosphere 
are the same. The simulated images are from a formation height of 
approximately 500~km whereas \citet{BeJ69} state that Ca~{\sc{ii}}~K is formed at a 
height less than 1300~km. Also, the presence of reverse granulation (as displayed in Figure~\ref{Fig1}) would 
suggest that the dominant contribution to the Ca~{\sc{ii}}~K filter comes from the upper photosphere/lower chromosphere. 
A review by \citet{Rut07} highlights the complexity of determining line formation regions from narrow band pass filters 
and suggests that Ca~{\sc{ii}}~K is formed at the 
lower chromospheric level. Although the simulated images 
do not have a formation height which is explicitly the same as the observed Ca~{\sc{ii}}~K core images, 
we think that a comparison can 
still be made. The relatively broad bandpass of the Ca~{\sc{ii}}~K filter make it difficult to assign a 
specific height of formation.  
However, there is a slight discrepancy between the values found at the 
formation height of the G-band, with the velocities of the observed data having an average of 
0.7~km\,s$^{-1}$ compared to 1.3~km\,s$^{-1}$ in the simulated images. This is possibly due to the far 
superior sample size used to ascertain this value in the simulated images. 
Both the simulated and observed datasets show that the average velocities of MBPs increase with height. 
This is probably a direct consequence of the conservation of momentum combined with a decrease in the density. 
However, it has been shown \citep{Jess10a} that the expansion of flux tubes as they extend upwards is affected by the 
field strength of the tube. Higher field strengths lead to less expansion of the flux tube at higher levels. 
This could also affect the velocities of the MBPs, but more studies would be required to ascertain what effect the 
field strength has, if any, on the estimated velocities. 

To understand  the nature of these flux tubes, and how they transfer energy to 
higher levels, we look for changes in the velocities of MBPs between different levels as a 
function of time, which may indicate how rigid 
these flux tubes are. This study was first performed on the simulations as they contained the velocity information for 
higher layers, making the analysis easier. From this analysis an average delay of 
 $\sim$4~s was found between the G-band simulations and a height of 500~km 
above the continuum formation layer where the velocity domain is taken.  
This may indicate that the flux tubes between the layers are not rigid. 
However, as the spread of values for the delays in velocity ranges from 
$\sim$$2-8$~s it suggests that the rigidity of these magnetic flux tubes can vary. 
Both the field strength and geometry of the individual tube can play a role in its flexibility. 


The average magnetic field strength of G-band MBPs in the simulations at the continuum formation level is about 950~G,
while that of the corresponding MBPs  detected at  500~km above the continuum formation layer is 450~G. 
For the MBPs studied, the magnetic field 
strength is therefore reduced by $\sim$50\% between the G-band images and the height in the atmosphere where we take 
the velocity measurements. Over the same range the density is reduced by approximately a factor of ten.

The short time delays in the simulated images at the continuum formation layer and a height of 500~km 
would imply that energy between the two layers is transported at a very high rate.   
Given that the sound and Alfv\'{e}n speeds are in the range $11-12$~km\,s$^{-1}$ and $10-15$~km\,s$^{-1}$, respectively, 
between the two layers, it seems unlikely that the observed delay is due to the movement of the footpoints in the simulation. 
A more likely scenario may be that the delays are produced by oblique granular shock waves, as reported by \citet{Vit11}.

A similar analysis was performed on the observed G-band and Ca~{\sc{ii}}~K datasets. The observations show 
an average delay of 29~s in the response of the velocity changes between the bandpasses, with the actual values ranging 
 between 21~s and 42~s.  

Due to the broad bandpass of the Ca~{\sc{ii}}~K filter, it encompasses radiation from both the 
chromospheric emission core and photospheric absorption profile.
It is therefore extremely difficult to assign a specific height from which this radiation 
originates. However, the reverse granulation present in the Ca~{\sc{ii}}~K images 
indicates that emission from the upper photosphere may dominate this bandpass. 
A rough estimate for the distance between the G-band and Ca~{\sc{ii}}~K emission region is around 500~km. Given a delay 
of 29~s, this suggests a velocity of propagation of $\sim$17~km\,s$^{-1}$ between the
two positions for the flux tubes. This would suggest that the propagation velocity is close to the sound speed, 
given typical values for the two positions. As the propagation 
speeds are slower in the observations, it suggests that the physical reason behind 
the delay may not be the result of granular shock waves. It is possible that the delay between 
the G-band and Ca~{\sc{ii}}~K observations is the result of two interwoven factors; the 
decrease in density and the expansion of the flux tube with height. The first factor is easily 
explained. With a decrease in the density of the surrounding material, the flux 
tube is less confined and therefore less responsive to the granular movements at the photospheric level, 
thus resulting in delays of velocity shifts. The second factor, which as stated previously 
is most likely a result of the decrease in 
density with height, probably has an added effect on the flexibility of the tubes. As the tube expands 
in the atmosphere, the magnetic flux contained within is less densely held in the tube and as a result 
it is less rigid. This may explain the difference in delays seen in the {\it{ROSA}} images. Unfortunately, we do not 
have magnetic field information for both G-band and Ca~{\sc{ii}}~K to assess how the field strength varies with height 
in the observations, and to investigate if there is any correlation between this and the delays in velocity shifts.

\section{Conclusions}
\label{conc}
We study the properties of MBPs as a function of height in the lower solar atmosphere 
using high resolution multi-wavelength imaging and 
radiative MHD simulations. The results show that the transverse velocity and area coverage 
of MBPs increases with 
height.  Velocities of 1~km\,s$^{-1}$ are observed for MBPs found in the G-band images, 
with the corresponding Ca\,{\sc{ii}}~K images yeilding an average velocity of 
1.4~km\,s$^{-1}$. The areas of the MBPs show an average increase of 
approximately 63\% from the 
photosphere to the chromosphere.  

Radiative MHD simulations reveal  similar findings but the theoretical 
values are slightly elevated when compared to the observations. 
Due to difficulties associated with simulating the radiative emission in 
Ca~{\sc{ii}}~K, we are unable to calculate changes in the area and velocity 
of MBPs in simulated Ca~{\sc{ii}}~K images. Instead we use the simulated 
horizontal velocity field and find that the delay in the changes of velocity
bewteen the two layers is approximately 4~s. 
We speculate that this delay may be attributed to horizontally-propagating 
oblique granular shock waves. 

The observations indicate an average delay of 29~s between the 
G-band and Ca~{\sc{ii}}~K. We surmise that the difference between the observations and the simulations is 
due to the height of formation of Ca~{\sc{ii}}~K, although the relatively broad bandpass of the Ca~{\sc{ii}}~K filter makes 
this height difficult to ascertain. It should also be noted that the difference in properties of MBPs in the observations 
and the simulations could be due to the different nature of turbulence conditions in the Sun and in the simulations, where 
the Reynolds number is very low. This is likely to affect the interaction of the flux tubes and convection at different altitudes.
We have presented one of the larger-scale studies of the properties of MBPs at multiple heights.
The results show that the transverse velocity of MBPs increase with height, most likely 
as a result of the decreasing density at higher levels and the conservation of momentum.

\section*{Acknowledgments}

This work has been supported by the UK Science and Technology
Facilities Council (STFC). Observations were obtained
at the National Solar Observatory, operated by the Association
of Universities for Research in Astronomy, Inc. (AURA), under
cooperative agreement with the National Science Foundation.
P.H.K. thanks the Northern Ireland Department for Employment
and Learning for a PhD studentship. D.B.J. thanks the STFC for
the award of a Post-Doctoral Fellowship. Effort sponsored by
the Air Force Office of Scientific Research, Air Force Material
Command, USAF under grant number FA8655-09-13085.

\end{document}